# Atomic motions in the $\alpha\beta$-region of glass-forming polymers: Molecular versus Mode Coupling Theory approach.

**Juan Colmenero**[1,2,3]§, **Arturo Narros**[1], **Fernando Alvarez**[1,2], **Arantxa Arbe**[2], **Angel J. Moreno**[3]

[1] Departamento de Física de Materiales UPV/EHU, Apartado 1072, 20080 San Sebastián, Spain
[2] Unidad Física de Materiales (CSIC–UPV/EHU), Apartado 1072, 20080 San Sebastián, Spain
[3] Donostia International Physics Center, Apartado 1072, 20080 San Sebastián, Spain

**Abstract.** We present fully atomistic Molecular Dynamics simulation results on a main-chain polymer, 1,4-Polybutadiene, in the merging region of the $\alpha$- and $\beta$-relaxations. A real space analysis reveals the occurrence of localized motions ("$\beta$-like") in addition to the diffusive structural relaxation. A molecular approach provides a direct connection between the local conformational changes reflected in the atomic motions and the secondary relaxations in this polymer. Such local processes occur just in the time window where the $\beta$-process of the Mode Coupling Theory is expected. We show that the application of this theory is still possible, and yields an unusually large value of the exponent parameter. This result might originate from the competition between two mechanisms for dynamic arrest: intermolecular packing and intramolecular barriers for local conformational changes ("$\beta$-like").

## 1. Introduction

Since they do not easily crystallize, polymers are probably the most extensively studied systems in relation with the glass-transition phenomenon and its associated structural relaxation. They are thus implicitly considered as "standard" glass-forming systems. However, the macromolecular character of their structural units must not be forgotten, as well as the chain connectivity, which plays an essential role in the dynamics of these systems. The most evident signature of this ingredient is the sublinear increase of the mean squared atomic displacements ("Rouse-like", $\langle r^2(t)\rangle \propto t^{0.5}$) arising after the decaging process ($\alpha$-process) in contraposition to the linear regime (center of mass diffusion, $\langle r^2(t)\rangle \propto t$) found in low-molecular weight glass-formers. Another peculiarity of polymers is that, apart from librations or fast rotations of methyl groups, every motion, as local as it be, involves jumps over carbon-carbon rotational barriers and/or conformational changes in the chain. Due to chain connectivity and intramolecular barriers, each conformational transition induces a perturbation that propagates along the chain backbone and influences the neighbouring atoms within the same macromolecule. It is easy to imagine that these systems are dynamically very rich –they indeed exhibit an enormous number of internal degrees of freedom.

§ To whom correspondence should be addressed (juan.colmenero@ehu.es)

1,4-Polybutadiene (1,4-PB) (-[$CH_2$-CH=CH-$CH_2$-]$_n$) has been considered as a "canonical" glass-forming polymer due to its lack of side groups. A series of experimental works have been focused on understanding the dynamics of this system above the glass-transition temperature $T_g$, in particular in the so-called $\alpha\beta$-merging region (in the neighbourhood of $T \approx 1.2T_g$). Such extensive studies have led to a puzzling situation: in the $\alpha\beta$-region, apart from the structural relaxation and the Johari-Goldstein $\beta$-relaxation observed by dielectric spectroscopy [1], at least other two faster processes are found by neutron [1] and light [2, 3] scattering, both being slower than the fast microscopic process below 1-2 ps [4]. The raised controversy demands a molecular understanding of the dynamics in this "simple" polymer, which can be facilitated by Molecular Dynamics (MD) simulations. With these ideas in mind, we performed fully atomistic MD-simulations on this polymer. The obtained results clearly showed the occurrence of localized motions in the $\alpha\beta$-regime, being active in the time window between the microscopic dynamics and the structural relaxation [5]. A recent molecular approach [5, 6] has identified such motions with the conformational transitions associated to the different secondary relaxations experimentally reported.

What is the impact of these motions on the application of the current theories for glass-forming dynamics to polymer data? Interestingly, we note that evidence of such local motions is found just in the time window where the Mode Coupling Theory (MCT) for the glass transition [7] predicts the occurrence of *its* $\beta$-process, which *a priori* is not related to the *Johari-Goldstein* $\beta$-process. Then, can the MCT be applied to 1,4-PB?

In this work we present MD-simulation results on 1,4-PB in the $\alpha\beta$-region. After summarizing the outcome of the recently developed molecular approach [5, 6], we address the question of the applicability of the MCT to this polymer. A rather satisfactory test of the MCT predictions is achieved –at least formally– though with an unusually large value of the exponent parameter $\lambda$. After comparing this result with data from the literature on a variety of systems of different nature, we speculate about the origin of this observation as the result of two competing mechanisms for dynamic arrest: intermolecular packing and intramolecular barriers for conformational changes.

## 2. Molecular Dynamics Simulations

Fully atomistic MD-simulations were carried out by using the Discover-3 module with the polymer consortium forcefield under the InsightII environment from Accelrys. A cubic cell containing one polymer chain of 130 monomers was constructed at 280 K (about 100 K above the experimental $T_g$) by means of the Amorphous Cell protocol with periodic boundary conditions. The microstructure of the chain (39% cis; 53% trans; 8% vinyl units) was built to mimic that of the real sample. After equilibration at 280

K the temperature was gradually lowered to the different investigated values (260 K, 240 K, 230 K, 220 K and 200 K) by a series of NPT dynamic steps at atmospheric pressure. For each temperature, once the equilibration density was reached, the energy of the structure was minimized and the system was dynamically equilibrated by a long run in the NVT ensemble. A subsequent NVT run was used to produce configurations for data analysis. Comparisons with extensive neutron scattering results showed that the simulated cell reproduces very fairly both structural and dynamical properties. A small shift in the simulation timescale provides an excellent agreement with experiments (simulation results are faster by less than half a decade). This shift roughly corresponds to a difference in temperature of 15 K. Further details can be found in Refs. [5, 6, 8].

## 3. Results

Figures 1(a) and (b) show the self-part of the radial distribution function, $4\pi r^2 G_s(r,t)$, with $G_s(r,t)$ the van Hove self-correlation function, averaged for all hydrogens in the cell at the highest and the lowest investigated temperatures. A qualitatively different behaviour can be found: at 280 K –well above $T_g$–, the distribution continuously broadens and shifts towards larger distances with increasing time. These are typical signatures of a diffusive process as that involved in the structural α-relaxation. Contrarily, at 200 K the most prominent feature in the distribution function is the development of a second peak located at $\approx 3$ Å. This double-peak structure reveals an underlying localized process as, e. g., a jump between two positions separated by a distance $d \approx 2.5$ Å (roughly the distance between both peaks). An inspection of $G_s(r,t)$ for each type of hydrogens in the sample reveals a strongly heterogeneous behaviour: the jumps occur in a distinct manner depending on the kind of atom considered. Finally, at this temperature, the characteristic features of an incipient diffusion can be only envisaged at the long-time limit of the simulation window.

The occurrence of these local motions strongly influences the short-time region of the correlation functions and other related quantities, as e. g. the mean squared displacement $\langle r^2(t) \rangle$ of the hydrogens [see Fig. 1(c)]. Again here a distinct behaviour is evident for high and low temperatures. At 280 K, the decaging mechanism takes place almost immediately after the microscopic regime (picosecond region). In contrast, at 200 K the slope $\approx 1/2$ (expected limit for the decaging process towards the Rouse regime) is reached at rather long times (beyond 10 ns). Interestingly enough, in the region between the microscopic and the final decaging dynamics, instead of the a priori expected plateau we find a continuous increase of $\langle r^2(t) \rangle$ (slope $\approx 0.2$). This feature suggests an additional dynamical process occurring within the cage imposed by the neighbouring molecules. Such a process is just that observed as a local motion in $G_s(r,t)$ [see Figs. 1(a) and (b)]. Moreover, at these low temperatures, such local processes also strongly affect the intermediate scattering function $F_s(Q,t)$ in the time regime between the microscopic dynamics and the structural relaxation, as can be appreciated from Fig. 1(d). A noticeable effect is observed even for low $Q$ values [5], where intuitively

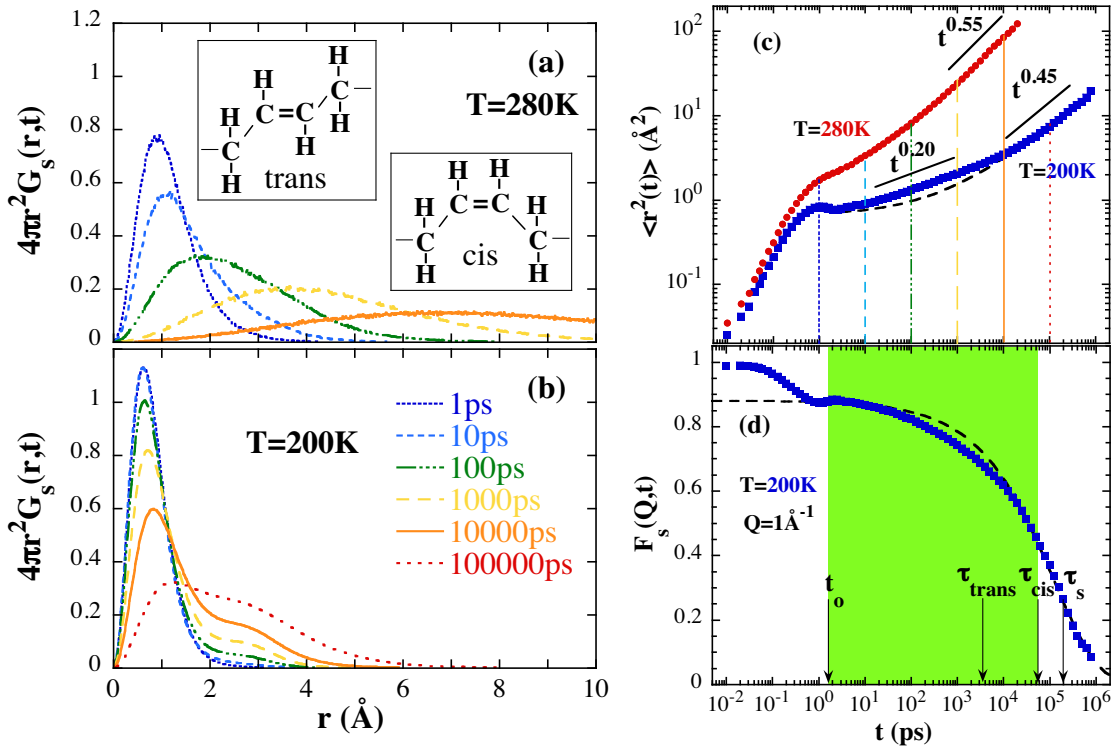

**Figure 1.** Simulations results for *all* the hydrogens. Panels (a) and (b): Radial distribution functions (self) at respectively 280 K and 200 K. Times are given in (b). Insets in (a): scheme of the cis and trans units. Panel (c): $\langle r^2(t)\rangle$ at 280 K (circles) and 200 K (squares). The vertical lines indicate the times considered in (a) and (b). Panel (d): $F_s(Q,t)$ at 200 K and $Q=1$ Å$^{-1}$. The thick dashed curves in (c) and (d) show the diffusive contribution from the jump-diffusion model (affected by the fast microscopic decay). Shadowed area: time interval most influenced by the localized processes. Vertical arrows indicate the characteristic time scales for the microscopic dynamics ($t_o$), for the jumps identified for the methyne hydrogens in the trans ($\tau_{trans}$) and cis ($\tau_{cis}$) units, and for the structural relaxation ($\tau_s$, KWW-time of the normalized dynamic structure factor at the first maximum of $S(Q)$, shown in Fig. 2(a)).

diffusion would be the dominant relaxation mechanism.

## 4. Molecular Approach

As shown in recent works [5, 6], the phenomenology observed for the hydrogen motions in the $\alpha\beta$-region of 1,4-PB can be well described by using a simple molecular approach which considers the simultaneous occurrence of localized motions and diffusion. In this framework, each atom jumps between two positions while it also participates in the anomalous diffusion related with the viscous flow associated with the structural relaxation. The obtained results can be summarized as follows: (i) the temperature dependence of the diffusion coefficient is consistent with the experimental expectation from viscosity measurements; (ii) the qualitative features deduced for this process are independent of the type of hydrogen considered; (iii) the activation energy deduced for the jumps of the methyne hydrogens (attached to the double bond) in the cis units is basically the same as that of the $\beta$-process monitored by dielectric spectroscopy [1] (we note that, consistently, the dipole moment in 1,4-PB is carried by the cis unit); (iv) the time scales obtained for the fastest hydrogens (methyne in the trans units) are in excellent agreement with those reported from Raman and Brillouin scattering [2, 3]. The jump distance involved in this case is well defined ($d_{trans} \approx 2.5$Å) and could be attributed to counterrotations of the trans units that leave the conformation of the rest of the chain practically unperturbed [9]. Other kinds of conformational transitions should be at the origin of the jump processes observed for the other hydrogens in the cell. This molecular approach thus provides a direct connection between the experimental observations of the different secondary relaxations reported for this polymer and the local motions involved in the conformational transitions taking place in the 1,4-PB building blocks.

We note that the region of maximum visibility of the localized motions is just located in the time window where one would expect the dynamics of the MCT $\beta$-process to be active. Thus, can the MCT be applied in this case and, if so, what is the result of a MCT analysis of our fully atomistic MD-simulations on 1,4-PB?.

## 5. Mode Coupling Theory Approach

The main concepts developed by the MCT are widely described, e. g., in Ref. [7]. Here we will just invoke some of the main predictions that will be tested with our data. As a first step towards a MCT analysis we may test the consistency of some of its predictions by considering the behaviour of parameters that can be easily obtained from a phenomenological data analysis. For example, the value of the von Schweidler exponent $b$ can be deduced as the high-$Q$ limit of the stretching parameter $\beta$ [10],

$$\lim_{Q\to\infty} \beta(Q) = b \, , \tag{1}$$

that results from the fit of the last stage of the decay of a given correlation function $\phi_Q(t)$ by a Kohlrausch-Williams-Watts (KWW) function

$$\phi_Q(t) = A\exp[-(t/\tau_{KWW})^\beta] \, . \tag{2}$$

As can be seen in Fig. 2(a), the high-$Q$ limit of $\beta$ is independent of the considered correlator and can be taken as $b \approx 0.24$, consistent with the range reported in experiments ($\beta : 0.17\ldots 0.40$ for $1.5 \leq Q \leq 5$Å$^{-1}$) [11]. This $b$-value is significantly lower than those usually found for low-molecular weight systems, as well as for simple bead-spring models for polymers (see below). Through the MCT relations

$$\lambda = \frac{\Gamma^2(1+b)}{\Gamma(1+2b)} = \frac{\Gamma^2(1-a)}{\Gamma(1-2a)}; \qquad \gamma = \frac{1}{2a} + \frac{1}{2b} \tag{3}$$

this value of $b$ implies unusually large values for $\lambda$ and $\gamma$ [$\lambda = 0.93$ and $\gamma = 4.9$]. Are these findings consistent with, e. g., the $T$-dependence of the characteristic time for the structural relaxation $\tau_\alpha$?. In the MCT, the way this time approaches the critical temperature $T_c$ is determined by the exponent $\gamma$ as

$$\tau_\alpha \propto |T - T_c|^{-\gamma} \ . \tag{4}$$

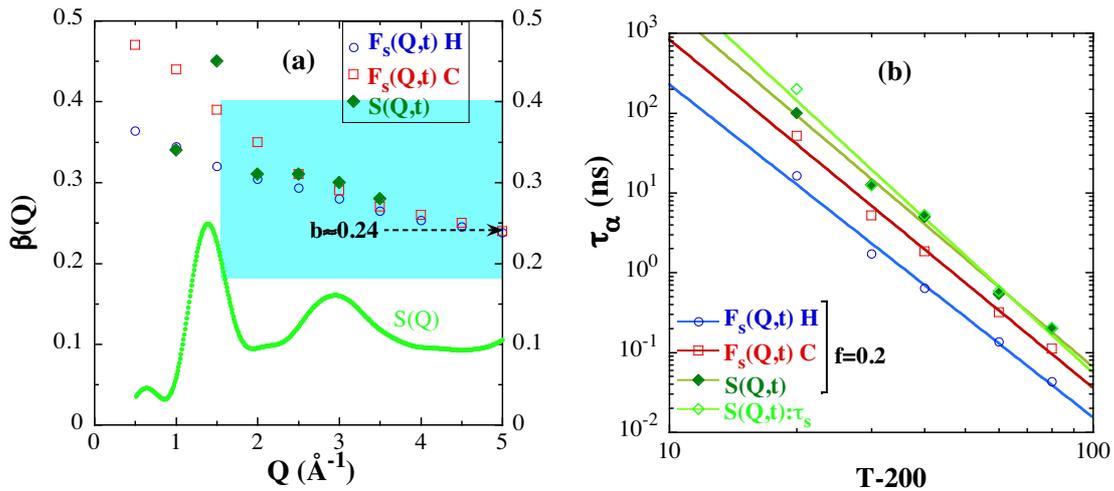

**Figure 2.** Panel (a): $Q$-dependence of the stretching parameter $\beta$ obtained for $F_s(Q,t)$ of all hydrogens (circles), of all carbons (squares), and for the collective dynamic structure factor $S(Q,t)$ (diamonds). Shadowed area: range deduced from neutron scattering on a fully protonated sample [11]. The static structure factor $S(Q)$ is shown for comparison in arbitrary units. (b) Temperature dependence of the characteristic time scales obtained from the same functions as in (a), defined as the time where the normalized correlation function takes the value $f = 0.2$. The selected $Q$-value corresponds to the maximum of $S(Q)$. For $S(Q,t)$ the KWW time is also considered. Lines: fits to power-laws with free exponents.

A priori, we do not know the value of $T_c$ for our system. An estimation could be $T_c \approx 1.2\, T_g$, as usually found in the literature. Taking into account the shift of the temperature in the MD-simulations below that in the real sample ($\Delta T \approx 15$K) and the experimental value of $T_g$ for 1,4-PB ($T_g \approx 180$ K), we find $T_c \approx 200$ K. Figure 2(b) shows that the fit to power laws corresponding to Eq. (4) with $T_c = 200$ K provide large exponents in the range $\gamma = 4.2\ldots 4.9$. As can be realized, these values are in very good agreement with the unusually large value of the $\gamma$-exponent independently deduced from Eqs. (1) and (3). Thus, from this first simple analysis we can conclude that: (i) independent tests of MCT predictions are consistent; (ii) the exponent parameter $\lambda$ seems to be unusually large, very close to 1.

Another prediction that can be tested without invoking fits is the factorization in the $\beta$-regime: $\phi_Q(t) = f_Q^c + h_Q\, G(t)$. If it holds, the ratio

$$R_Q(t) = \frac{\phi_Q(t) - \phi_Q(t')}{\phi_Q(t'') - \phi_Q(t')} \tag{5}$$

must be $Q$-independent. As shown in Fig. 3 for $F_s(Q,t)$ at 200 K, the superposition of these functions is almost perfect in the MCT-$\beta$ region (2 ps $\lesssim t \lesssim$ 1 ns) –i.e., in the time window where the influence of the localized motions is most evident!

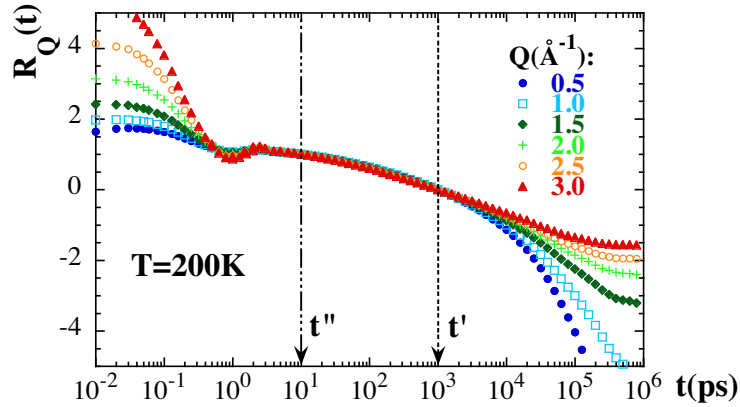

**Figure 3.** Test of the factorization theorem for $F_s(Q,t)$ of all hydrogens at 200 K.

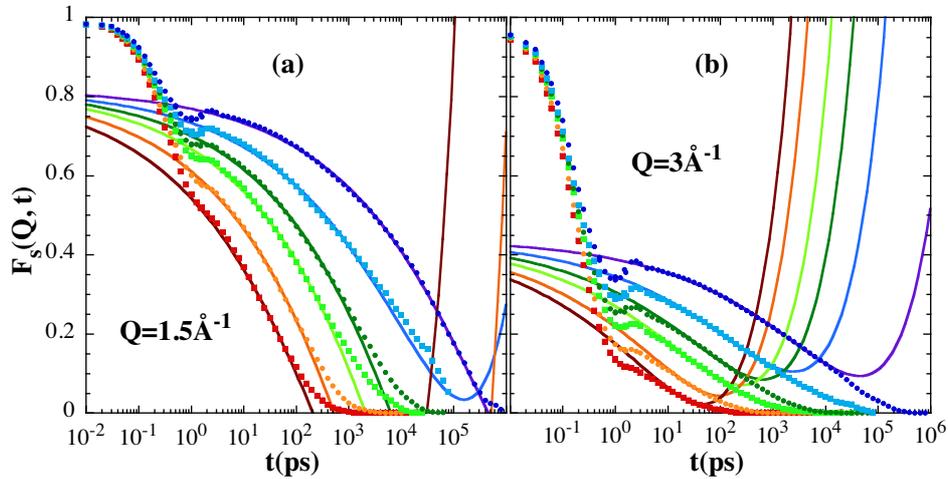

**Figure 4.** Fit of the MCT-$\beta$ regime for $F_s(Q,t)$ of all hydrogens at $Q = 1.5$ Å$^{-1}$ (a) and $Q = 3$ Å$^{-1}$ (b) and 200, 220, 230, 240, 260 and 280 K (top to bottom).

From now on we will thus consider that the MCT can be applied to our data with the critical exponents determined from the previous analysis. Being the $\beta$-scaling fulfilled, a description of the $\beta$-regime by means of the MCT von Schweidler expansion

$$\phi_Q(t) = f_Q - H_{1Q}t^b + H_{2Q}t^{2b} + \ldots \qquad (6)$$

with a $T$-independent critical non-ergodicity parameter $f_Q$ should be possible. We have first considered $F_s(Q,t)$ of all hydrogens at different temperatures. For each $Q$-value, we have described all the curves (Fig. 4) with a single value of $f_Q$. The resulting $f_Q$-values are displayed in Fig. 5, together with those obtained from the fit to Eq. (6) of the normalized collective dynamic structure factor $S(Q,t)/S(Q)$ at two temperatures. Again for this correlator $f_Q$ does not depend on temperature. Moreover, it is modulated by $S(Q)$. Both observations agree with MCT expectations.

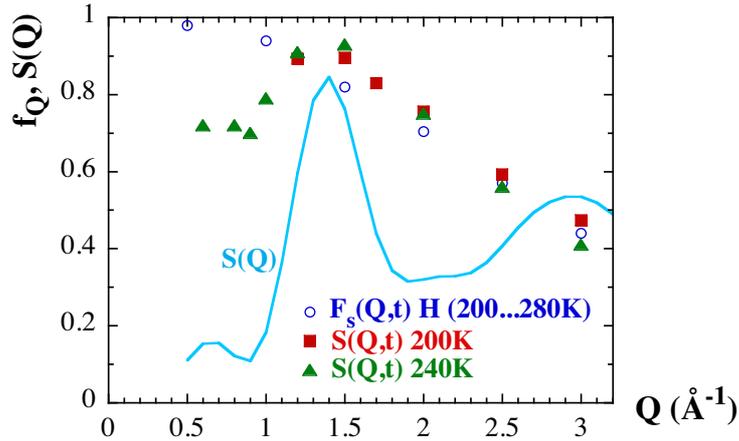

**Figure 5.** $Q$-dependence of the critical non-ergodicity parameter obtained from $F_s(Q,t)$ of all hydrogens (circles) and from $S(Q,t)/S(Q)$ at 200 K (squares) and 240 K (triangles). The static structure factor is shown for comparison.

Finally we can obtain the value of the critical temperature $T_c$ of our sample. This can be realized in two ways: (i) from the MCT description in the von Schweidler regime, since the $H_{1Q}$ parameters in Eq. (6) must fulfill the relation

$$H_{1Q} \propto |T - T_c|^{\gamma b} ; \qquad (7)$$

(ii) from the time scale of the $\alpha$-regime [Eq. (4)]. Approach (i) leads to $T_c = 188$ K [see Fig. 6(a)], while through approach (ii) we arrive at $T_c = 198$ K [Fig. 6(b)]. Both values, differing by only a 5%, are compatible within error bars. It is worthy of remark that the range of values obtained for $T_c$ is also compatible with the $T_c$-value reported from experimental studies on this polymer ($T_c \approx 214$ K) [12], taking into account the above mentioned shift in temperature between the simulated and real samples.

## 6. Discussion

We have shown that atomic motions in the $\alpha\beta$-regime ($200 \leq T \leq 280$ K) of 1,4-PB consist of localized and diffusive contributions. From a molecular approach, the localized processes can be associated with conformational transitions involving the rigid building blocks of 1,4-PB that are responsible for the experimentally observed secondary relaxations (including the Johari-Goldstein $\beta$-relaxation). These motions strongly affect the dynamics between the microscopic region and the structural relaxation, i. e., in the MCT $\beta$-regime dominated by the "cage" effect imposed by the surrounding neighbours. However, at least formally, MCT can be applied to 1,4-PB results and its main predictions are fulfilled. Apparently, the MCT analysis reflects the presence of the local conformational changes through the large value deduced for the exponent parameter $\lambda$. As mentioned above and shown in Table 1, MCT analysis of different systems [13, 14, 15, 16] yield much smaller $\lambda$-values. Low-molecular weight systems, as o-terphenyl, silica or water, show values similar to that of hard spheres. However, adding the ingredient of chain connectivity in bead-spring models does not yield significant changes in $\lambda$. We note that bead-spring models referred to in [17] deal with *fully flexible* chains and do not include rotational barriers. A significant change is induced by incorporating these barriers, as in an united atom model of polyethylene [18]. Even larger $\lambda$-values are obtained for 1,4-PB. Seemingly, the presence of double bonds along the main chain of 1,4-PB allows for counterrotations leading to particular local motions of parts of the macromolecule which are not found in the most simple polyethylene. Our study suggests that the role of the intramolecular barriers is crucial in determining the dynamics within the cage and, as a consequence, the observed large value of $\lambda$.

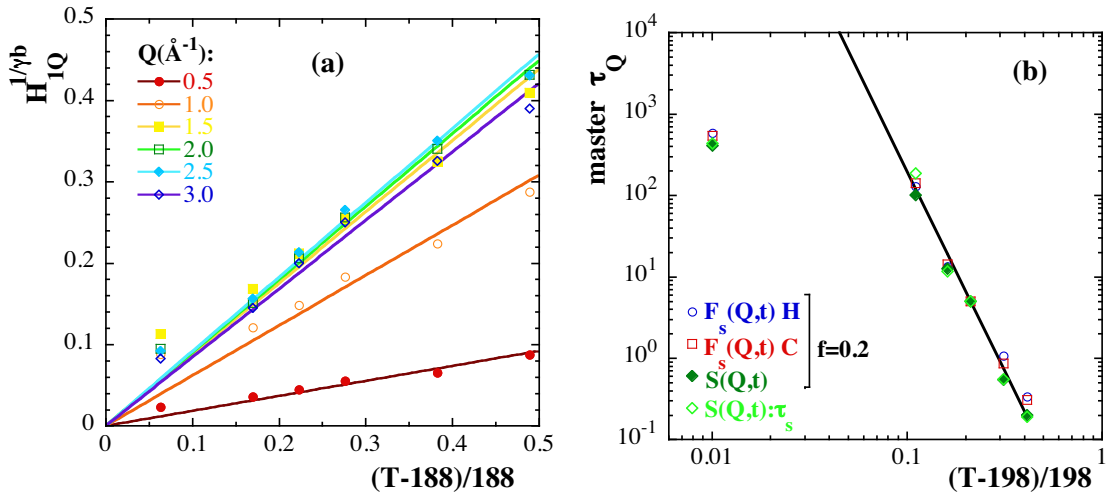

**Figure 6.** Tests of MCT predictions. Panel (a): $\beta$-scaling [Eq. (7)]; panel (b): $\alpha$-relaxation times defined as in Fig. 2(b) and scaled into a master curve [Eq. (4)]. As usually observed, deviations from MCT predictions, which are attributed to ill-defined ergodicity-restoring processes, occur at temperatures very close to $T_c$.

**Table 1.** Values of the MCT critical exponents for different systems

| System | $a$ | $b$ | $\gamma$ | $\lambda$ | Reference |
|---|---|---|---|---|---|
| Hard spheres | 0.31 | 0.58 | 2.5 | 0.74 | [13] |
| o-terphenyl | 0.30 | 0.54 | 2.6 | 0.76 | [14] |
| Silica | 0.32 | 0.62 | 2.3 | 0.71 | [15] |
| Water | 0.29 | 0.51 | 2.7 | 0.78 | [16] |
| Bead-spring polymer (MD) | 0.35 | 0.75 | 2.1 | 0.63 | [17] |
| Bead-spring polymer (MCT) | 0.32 | 0.60 | 2.4 | 0.72 | [17] |
| Polyethylene (united atom) | 0.27 | 0.46 | 2.9 | 0.81 | [18] |
| Polybutadiene (united atom) | 0.21 | 0.30 | 4.1 | 0.90 | [19] |
| Polybutadiene (fully atomistic) | 0.18 | 0.24 | 4.9 | 0.93 | This work |

Values of $\lambda$ approaching 1 point to a situation close to a MCT higher-order transition ($\lambda = 1$) [20]. Features associated to such transitions have been recently reported for attractive colloids [21] and a series of systems showing strong confinement effects [22]. The origin of the observed anomalous relaxation features is attributed to the presence of several competing mechanisms for dynamic arrest –steric repulsion and reversible bond formation in [21], bulk-like caging and confinement in [22]. In the case of the $\alpha\beta$-region of real polymers we may speculate that there also exist two active competing mechanisms leading to dynamic arrest: i) packing, of intermolecular character and present in all glass-forming systems; ii) barriers for conformational changes ("$\beta$-like"), of intramolecular origin, which are specific of macromolecular systems.

## 7. Acknowledgments

We acknowledge support from the European Commission NoE SoftComp, Contract NMP3-CT-2004-502235, the projects MAT2004-01017 and 9/UPV00206.215-13568/2001, and from Donostia International Physics Center. A. N. acknowledges a FPI grant of the Spanish Ministery of Science and Technology.